 \documentclass[nohyper,12pt,letterpaper]{JHEP3}

 \title{Holographic space-time, cosmological SUSY breaking, and particle phenomenology}
 \author{T.\,Banks\\
 Department of Physics and SCIPP\\
 University of California, Santa Cruz, CA 95064\\
 E-mail: \email{banks@scipp.ucsc.edu}\\
 {\it and}\\
 Department of Physics and NHETC, Rutgers University\\
 Piscataway, NJ 08540 }

 \abstract{I briefly review the theory of Holographic Space-time and its relation to the
cosmological constant problem, and the breaking of supersymmetry (SUSY). When
combined with some simple phenomenological requirements, these ideas lead to
a fairly unique model for Tera-scale physics, which implies direct gauge mediation
of SUSY breaking and a model for dark matter as a hidden sector baryon, with nonzero
magnetic dipole moment.
 }
 \received{} \accepted{}
 \preprint{hep-th{}\\\\ \\}
 
\begin{document}
\section{Introduction to holographic space-time\cite{holost}}

This paper is the written version of a talk given at the conference celebrating the 80th birthday
of Murray GellMann at the Nanyang Technical University in Singapore. I'd like to thank Harald Fritzsch and
the other organizers of the conference for inviting me to join in honoring one
of the greatest physicists of the 20th century.

String theory models are our only rigorously established models of quantum
gravity, but none of the known models apply to the real world. They do not
incorporate cosmology, and they do not explain the breaking of supersymmetry
(SUSY) that we observe.
The theory of Holographic Space-Time is an attempt to generalize string
theory in order to resolve these problems. Its basic premise is a strong
form of the holographic principle, formulated by myself and W. Fischler:
\begin{itemize} \item Each causal diamond in a $d$ dimensional Lorentzian
space-time has a maximal area space-like $d - 2$ surface in a foliation of
its boundary. The area of this {\it holographic screen} in Planck units is
$4$ times the logarithm of the dimension of the Hilbert space describing all
possible measurements within the diamond. \end{itemize}

Every pair of causal diamonds has a maximal area causal diamond in their
intersection. This is identified with a common tensor factor in the Hilbert
spaces of the individual diamonds
$${\cal H}_1 = {\cal O}_{12} \otimes {\cal N}_1$$
$${\cal H}_2 = {\cal O}_{12} \otimes {\cal N}_2 .$$ A holographic space-time
is defined by starting from a $d-1$ dimensional spatial lattice, which
specifies the {\it topology of a particular space-like slice}.  To each
point ${\bf x}$ on this lattice, we associate a sequence of Hilbert spaces
      $${\cal H} (n, {\bf x}) = \otimes {\cal P}^n .$$
The {\it single pixel Hilbert space}, ${\cal P}$ will be specified below,
and has to do with the geometry of compactified dimensions. These spaces
represent the sequence of causal diamonds of a time-like observer as the
proper time separation of its future tip from the point where it crosses the
space-like slice increases. $N ({\bf x})$ is the maximal value that $n$
attains as the proper time goes to infinity. In a future asympotically dS
space time,  $N({\bf x})$ will be finite.  In an asymptotically flat
space-time or FRW universe which is matter or radiation dominated,$n$ will
go to infinity with the proper time, while in an asymptotically AdS universe
$n$ will go to infinity at finite proper time.  In a Big Bang space-time the
past tip of each causal diamond lies on the Big Bang hypersurface. In a time
symmetric space-time we think of the diamonds as having past and future tips
which are equidistant in proper time from the slice on which the lattice is
placed. In either case we will refer to {\it the causal diamonds at a fixed time} as those
carrying the same label $n$.

In any theory of quantum gravity, the Hilbert space formulation will refer to
a particular time slicing.  We have chosen slices such that the causal diamonds at
any fixed time have equal area holographic screens. Such equal area slicings exist
in all commonly discussed classical space-times.

The rest of the specification of holographic space-time consists of a
prescription of the overlap tensor factor $${\cal O} (m , {\bf x}; n, {\bf
y}),$$ in ${\cal H} (m, {\bf x}) $ and ${\cal H} (n, {\bf y})$. For nearest
neighbor points at $m = n$ this overlap is just ${\cal P}$. For other pairs
of points the specification of the overlap is part of the dynamical
consistency condition described below. The only kinematic restriction on it
is that the dimension of ${\cal O}$ is a non-increasing function of $d({\bf
x,y})$, the minimum number of lattice steps between the points.

We introduce dynamics as a sequence of unitary operators $U(n, {\bf x})$ in
${\cal H} (N({\bf x}), {\bf x}) $, with the property that $U(n, {\bf x}) =
V(n, {\bf x}) W(n, {\bf x})$, where $V(n)$ is a unitary in ${\cal H} (n,
{\bf x})$, while $W(n)$ is a unitary in the tensor complement of ${\cal H}
(n, {\bf x})$ in  ${\cal H} (N({\bf x}), {\bf x}) $. This requirement
implements the idea that the dynamics inside a causal diamond effects only
those degrees of freedom associated with the diamond.  In particular, in a
Big Bang space-time, it builds the concept of {\it particle horizon} into the
dynamics of the system.  Note by the way that in Big Bang space time the
sequence of unitaries $U(n)$ may be thought of as a conventional time dependent
Hamiltonian system with a discrete time, while for a time symmetric space-time
they are instead ``approximate S-matrices", $U(T, - T)$.

Starting from some initial pure state in ${\cal H} (N({\bf x}), {\bf x})$, the
unitaries $U(n, {\bf x})$ produce a sequence of density matrices $\rho (n, {\bf x})$ in
each overlap factor involving the point ${\bf x}$. {\it The key dynamical consistency
condition for a holographic space time is that
$$\rho (n, {\bf y}) = U(n, {\bf x};  {\bf y}) \rho (n, {\bf x}) U^{\dagger} (n, {\bf x};  {\bf y})
,$$ for every pair of points.} This staggeringly complicated set of consistency
conditions is the analog in this formalism of the Dirac-Schwinger commutation
relations, which guarantee the consistency of ``many fingered time". The only known
solution of these conditions is the dense black hole fluid (DBHF) cosmology
described briefly below. In that example, the consistency conditions dictate both
the choice of overlap Hilbert spaces, and the dynamics at each point in the
lattice.

There are a number of very important points to understand about this formalism
\begin{itemize}

\item Although we have used geometrical pictures to motivate our constructions,
they are entirely phrased in quantum mechanical language. The Lorentzian space-time
is an emergent property of these quantum systems, useful in the limit of large
causal diamonds (large dimension Hilbert spaces).

\item The emergent space-time geometry is {\it not} a fluctuating quantum variable.
Its causal structure is specified by the overlaps, and its conformal factor by
the Hilbert space dimensions.

\item The lattice specifies only the topology of a space-like slice in the non-compact
dimensions\footnote{In a holographic theory, dS space has non-compact spatial
sections because one restricts attention to the causal diamond of a fixed observer.}
This topology does not change with time.

\end{itemize}

\section{SUSY and the holographic screens\cite{susyholo}}

Since space-time geometry is {\it not} a fluctuating quantum variable, it is natural to associate the
quantum variables with the properties of the holographic screen of the causal diamond. Intuitively, the
space-time orientation of an infinitesimal bit of screen is determined by the outgoing null direction, and the transverse
plane in which the screen lies.  That information is encoded in the Cartan-Penrose equation

$$\bar{\psi} \gamma^{\mu} \psi (\gamma_{\mu})^{\alpha}_{\beta} \psi^{\beta} = 0.$$ Indeed this equation implies that
$\bar{\psi} \gamma^{\mu} \psi$ is a null vector, and that the hyperplanes
$$\bar{\psi} \gamma^{\mu_1 \ldots \mu_k} \psi,$$ with $k \geq 2$ all lie in a single $d-2$ plane.
More succinctly, $\psi = (0, S_a)$: $\psi$ is a transverse spinor in the light front frame defined by the null direction.

The Cartan-Penrose equation is conformally invariant, but our quantization procedure will violate that invariance.
This is simply the statement that the Bekenstein-Hawking area formula is being used to define the conformal factor of
our space-time geometry in terms of the dimension of the quantum Hilbert space. The holographic principle
now implies two constraints on the quantization procedure:

We want to have independent degrees of freedom for different points on the holographic screen. This is compatible with a finite
dimensional Hilbert space only if a finite area screen is ``pixelated": its function algebra must be replaced by
a finite dimensional algebra. If $n$ labels a basis of the algebra, the single pixel Hilbert space ${\cal P}$ is the lowest dimension
representation space of the algebra generated by the $S_a (n)$ variables.  If we insist on transverse $SO(d-2)$ invariance, the only quantization
rule having a finite dimensional representation space is
$$[S_a (n), S_b (n) ]_+ = \delta_{ab}$$ SUSY aficionadas will recognize this as the algebra of a single massless supersymmetric particle with longitudinal momentum
proportional to $(1, \overrightarrow{\Omega})$, where $\Omega$ is the angular position of the pixel on $S^{d-2}$.
If $d=11$ the smallest representation of this algebra is the SUGRA multiplet. In fewer non-compact dimensions there are non-gravitational multiplets,
but, since we are trying to construct a theory of gravity, we should retain $16$ real spinor generators for each pixel, in order to guarantee that there is a helicity two particle in the
spectrum.

The anti-commutation relations postulated so far are invariant under $S_a (n) \rightarrow (-1)^{F(n)} S_a (n)$.
This is a remnant of the rescaling symmetry of the CP equation. We treat it as a gauge symmetry. Using it,
we can perform a Klein transformation so that the independent operators on different pixels anti-commute rather than
commute with each other.

A convenient way to pixelate the holographic screen is to use fuzzy geometry.  We replace the algebra of functions by a sequence of finite dimensional matrix algebras.
The most famous example is the two sphere.  The algebra of $n\times n$ matrices has a natural action of the group $SU(2)$ on it because the spin $\frac{n-1}{2}$ representation is
$n$ dimensional.  The matrices carry every spin from zero up to $n-1$ and so can be thought of as a natural cutoff of the angular momentum on the sphere. Vector bundles over the sphere
are rectangular matrices.  In particular $n \times n+1$ and $n+1 \times n$ matrices converge to the two chiral spinor bundles over the sphere. Many of the compactification spaces of string
theory are Kahler manifolds, or Kahler fibrations over a one (Horava-Witten) or three dimensional (G2 manifolds which are K3 fibrations) base.  These are naturally thought of as limits of finite
dimensional matrix algebras. The pixel variables of such compactifications will have the quantum algebra

$$ [ (\psi^M )_i^A , (\psi^{\dagger N})^j_B ]_+ = \delta_i^j \delta^A_B B^{MN} ,$$ where $i,j = 1 \ldots K$ and $A,B = 1 \ldots K + 1$ , so that the fermionic matrices fill out the two spinor bundles
over the fuzzy two sphere.  The indices $M,N$ also run over a set of rectangular matrices which approximate either the spinor bundle over a seven manifold, or two copies of the spinor bundle
over a six manifold (and there are two possibilities, according to whether the two copies have the same or different chiralities).   The $B^{MN}$ should be interpreted as wrapped brane charges.
They can be further decomposed into sums over cycles of various dimensions.  That is to say, we have an algebraic way of encoding the homology of the manifold\footnote{Though of course, we know from
string duality that the interpretation as homology of a particular manifold will only be valid in certain limits. The algebra of SUSY charges (of which our algebra is an analog) is valid
independently of the geometric interpretation.}.  In this formalism, the problem of (kinematically) classifying four dimensional compactifications reduces to classifying superalgebras such that
in the limit, $K \rightarrow \infty$ they contain one copy of the $N=1$ SUSY algebra.  Equivalently, in this limit, the representation space of the algebra should contain exactly one $N=1$ graviton
supermultiplet.  The super-generators are constructed by using the conformal structure of the 2-sphere, whose invariance group is $SO(1,3)$. Conformal Killing spinors on the sphere transform as the
Dirac spinor of $SO(1,3)$.

If $(q_{\alpha})^i_A$ are matrices that converge to the left handed conformal Killing spinor, then our kinematic condition on the algebra of pixel variables is that there exists a set of coefficients $F_M$
such that $$Q_{\alpha} \equiv F_M {\rm Tr}\ [\psi^M q_{\alpha} ] ,$$ satisfies the super-Poincare algebra as $K \rightarrow\infty$.   The representation space of the pixel algebra should break up
into a finite number of single particle representations of the SUSY algebra, with only a single supergraviton multiplet. This is, in our formalism, the condition for a compactification with $N = 1$ SUSY.
Note that, for finite $K$, these constructions have no continuous moduli.  They are finite dimensional unitary representations of finite dimensional non-abelian super-algebras.

\subsection{Particles}

If we suppose that we have found such an algebra, we can now make multiple copies of our single particle Hilbert space by replacing the algebra of functions, by the matrix algebra
$${\cal M}_K \otimes {\cal A}, $$ (where ${\cal A}$ is the algebra of matrices approximating the function algebra on the internal manifold), by a direct sum
$$\oplus_i {\cal M}_{Ki} \otimes {\cal A} , $$ and take the limit $K_i \rightarrow\infty$ with $\frac{K_i}{K_j}$ fixed.  As in Matrix Theory\cite{bfss} the ratios are interpreted as the ratios of
longitudinal momenta $P_i (1, {\bf \Omega_i})$ of a set of particles.  Here however, each particle has its own null direction.
The $S_p$ gauge symmetry relating commuting operators to block diagonal matrices is interpreted as particle statistics.  Note that a particle must have a large momentum in order to have
good angular localization, but for fixed holographic screen area one can only make a finite number of particles, and the larger the momentum that each one carries the fewer particles we can make.
One can argue\cite{bfm} that the states with all the momentum carried by one ``particle" should actually be thought of as black holes that
fill the causal diamond.

\subsection{Holographic cosmology}

Here we give a brief description of the Dense Black Hole Fluid model of holographic cosmology\cite{holocosm}. In this model,
one takes the overlap Hilbert spaces to be

$${\cal O} (n, {\bf x}; n {\bf y}) = {\cal P}^{n - d({\bf x , y})}, $$ where $d({\bf x , y})$ is the minimum number of lattice steps between the two points.  If the exponent is negative, we interpret it
as $0$.   The time evolution operators are identical at each lattice point, and the time dependent Hamiltonian is chosen randomly at each time $n$ to be
$${\rm ln}\ V(n, {\bf x}) = \sum S_a (i) S_a (i) A(n; i, j) + I (n) .$$ Here the $S_a$ satisfy fermionic commutation relations\footnote{We have not yet made a cosmology compatible with
the more complicated superalgebras that arise for non-trivial compactifications.}, and $A(n; i,j)$ is a random $n \times n$ anti-symmetric matrix. For large $n$ the quadratic term converges to
the Hamiltonian for free massless fermions in $1+1$ dimensions, and $I(n)$ is chosen to be a random irrelevant perturbation of this CFT.   One then argues that there is a coarse grained description of
this system as a flat FRW universe, with equation of state $p = \rho$, which saturates the covariant entropy bound.

This model is used to construct more realistic cosmologies by using the Israel junction condition.  We think of our own universe as a low entropy ``defect" inside the DBHF. Consider first
a spherical volume of $p = w \rho$ universe with $-1 < w < 1$, embedded in a $p=\rho$ universe. Consider time slices of the two geometries of equal holographic area.  This means the time coordinates are
proportional to each other with a fixed constant.
A coordinate volume of radius $L$ has a physical radius that grows as $t^{\frac{2}{3(1+w)}}$. Since the physical radius in the DBHF grows more slowly, we must let the coordinate $L$ shrink with time
in the $p = w \rho$ universe in order to satisfy the Israel condition that the geometry of the interface be the same in both embeddings. The exception is $w = -1$.  In this case, a cosmological
horizon volume is bounded by a null surface of fixed holographic area.  We can satisfy the Israel condition by matching to a black hole with the same area horizon, embedded in the $p = \rho$ background.

In \cite{holocosm} we argued that non-spherical defects could survive as  $ - 1 < w < 1$ regions, but that the above
argument about the Israel condition implies that eventually the universe must approach $w = -1$.  The late time cosmological constant is
determined by cosmological initial conditions, namely the number of degrees of freedom that are initially in a low entropy state. In this way of realizing dS space,
it is clear that only a single horizon volume of the classical geometry is necessary to the description, and that this is described as a quantum system with a finite
number of states: the representation space for the pixel algebra over the finite area holographic screen of the cosmological horizon.

\section{Cosmological SUSY breaking\cite{susyholo}\cite{bfm}\cite{holost}}

We have noted that in holographic cosmology, the cosmological
constant $\Lambda$ is a positive tunable parameter, determined by
cosmological initial conditions. To discuss particle physics, we can
replace the actual cosmological history with that of an eternal dS
space. In the limit $\Lambda \rightarrow 0$ the theory of stable dS
space approaches a super-Poincare invariant theory, similar to
conventional string theories, but with no moduli. The theory has a
discrete R symmetry, explaining the vanishing of the superpotential
at the supersymmetric point. However, the way in which this limit is
approached is interesting. One horizon volume of dS space approaches
all of Minkowski space. The logarithm of the total number of quantum
states of the dS theory is $\pi (RM_P)^2$, but only $~ (RM_P)^{3/2}
$ of that entropy can be modeled by field theory in the horizon
volume.

This entropy bound can be derived in two complementary ways\cite{nightmare}\cite{bfm}.
On the one hand, we can try to maximize the particle entropy in a horizon volume, subject to
the constraint that no black holes of size that scales like $R$ are formed. The maximal entropy particle states
are modeled  as a cutoff CFT with cutoff $\mu$, so that the entropy,
$$S\sim (\mu R)^3 .$$ The condition that no horizon sized black holes are formed is
$$ \mu^4 R^3 < M_P^2 R ,$$ which leads to $\mu < (M_P / R)^{1/2}$, and the $(R M_P)^{3/2}$ scaling
of the particle entropy.   On the other hand, if we model the holographic screen of dS space as a fuzzy sphere
with $K \propto R M_P $, and particle states by block diagonal fuzzy spheres of block sizes $K_i$ with $\sum K_i = K $, then the complementary
constraints of angular localization (maximizing each $K_i$) and maximizing the multiparticle particle entropy, lead to $K_i \sim \sqrt{K}$. If our basic
unit of longitudinal momentum is $1/R$, then this gives the same scaling for entropy, momentum cutoff, and average particle number as the previous argument.
These remarks also lead to a conjecture for what the other, off diagonal bands, of the matrices represent. The total entropy of dS space
allows us to have $(R M_P)^{1/2}$ independent copies of the field theoretic degrees of freedom in a single horizon volume, and it is an obvious conjecture that this is the
way the classical geometric result that at late global times dS space has an unbounded number of independent horizon volumes, is realized in the limit $RM_P \rightarrow\infty$.
The off block diagonal bands of the $K\times K$ matrix algebra approximation to the function algebra on the 2-sphere, represent the particle degrees of freedom in
different horizon volumes.

It is the field theoretic states in a single horizon volume, which approach the
scattering states of the Minkowski theory. The exponentially
overwhelming majority of the states of the dS theory decouple in
this limit\footnote{In quantum field theory, this is the statement
that the dS temperature goes to zero.}.  These states should be
viewed as living on the cosmological horizon. However, because their
number is so large, the effect of the interaction of localized
particles with the horizon states may be larger than one might have
imagined.

The discrete R symmetry of the $\Lambda = 0$ theory is broken by
interactions with the horizon. The lightest particle in the theory
carrying R charge is the gravitino. Thus, R violating interactions
will be dominated by Feynman diagrams in which a gravitino
propagates out to the horizon.  These are
suppressed by a factor $e^{- 2 m_{3/2} R}$. The contribution from
the interaction with the horizon has the form
$$ \sum_n \frac{|< \tilde{g} | V | n >|^2}{\Delta E} . $$ Note that there is no $n$ dependence
energy denominators in this formula, because the horizon states are
approximately degenerate.  To estimate the number of states that
contributes to this formula we note that the horizon states, like
degenerate Landau levels, can be localized and have a fixed entropy
per unit area. The gravitino can propagate in the vicinity of the
horizon, a null surface, for a proper time of order
$\frac{1}{m_{3/2}}$. Quantum particles execute random walks in
proper time. If we take the step size to be Planck scale, the area covered will also scale like
$\frac{1}{m_{3/2} M_P}$. Thus, the contribution of this diagram is of
order $$e^{-2 m_{3/2} R + \frac{b M_P}{m_{3/2}}},$$ where $b$ is an
unknown constant. We know that $m_{3/2} \rightarrow 0$ as $R
\rightarrow\infty$. If it went to zero faster than $R^{-
\frac{1}{2}}$, then the diagram would blow up exponentially. If it
goes to zero more slowly than $R^{- \frac{1}{2}}$, then the diagram
is exponentially small.  However, it is precisely the R violating
terms in the effective Lagrangian, which are supposed to be
responsible for the non-zero gravitino mass.  So we have a
contradiction unless $m_{3/2} = K \Lambda^{1/4}$.  In
\cite{pyramid1} we gave an argument that the constant $K$ is of
order $10$.

\section{CSB and phenomenology\cite{pyrma}}

The relationship $m_{3/2} = K \Lambda^{1/4}$, with $K$ of order
$10$, puts strong constraints on low energy phenomenological models.
In low energy effective SUGRA models, SUSY breaking is parametrized
by a non-vanishing F term for some chiral superfield, $X$. In order
to obtain gaugino masses, the model must generate couplings of the
form
$$ c_i \frac{\alpha_i}{4\pi} \frac{X}{M} {\rm tr} (W_{\alpha}^i)^2
.$$  Since, according to CSB, $F_X = K ({\rm TeV})^2 $, we cannot
have $M$ larger than a few TeV. Thus we {\it must} have a strongly
coupled hidden sector to generate the scale $M$, and that sector
must contain particles charged under the standard model gauge group.
That is, we have a model of direct gauge mediation.

If we wish to preserve the prediction of SUSY gauge coupling
unification, the new particles must be in complete multiplets of a
unified group, and transform under the hidden sector group $G$. If
the unified group contains $SU(5)$ we get, at least $R$ copies of
the $5 + \bar{5}$, where $R$ is a $G$ representation. If $R \geq 5$,
this leads to Landau poles below the unification scale, which
implies at best a fuzzy prediction of unification. All hidden sector
groups with $R < 5$ appear to predict light pseudo-Goldstone bosons
that transform under the standard model and should have been seen in
experiments.

The only resolution I have found to these competing exigencies is to
employ {\it trinification}\cite{trin}, with a hidden sector group
$SU_P (3)$.  The resulting model has a pyramidal quiver diagram and
is called The Pyramid Scheme. It has perturbative one loop
unification, and no unpleasant PNGBs. The gauge group is $SU_P (3)
\otimes SU_1 (3) \otimes SU_2 (3) \otimes SU_3 (3) \rtimes Z_3$, and
the matter content is $$3 \times [(1,1,\bar{3},3) \oplus
(1,3,1,\bar{3}) \oplus (1,\bar{3}, 3, 1)],$$ $$ (3,\bar{3},1,1)
\oplus (3,1,\bar{3},1) \oplus (3,1,1,\bar{3}) \oplus c.c. $$ The
$Z_3$ symmetry permutes the last 3 $SU(3)$ subgroups. The
$SU(2)\times SU(3)$ of the standard model is embedded in the
indicated $SU_{2,3} (3)$ groups of the Pyramid, and the $U(1)$ is a
combination of a generator in $SU_1 (3)$ and one in $SU_2 (3)$. In
addition, we introduce 3 singlet fields $S_i$.  The three fields
that couple both to $SU_P (3)$ and to the standard model are called
{\it trianons} and are denoted $T_i + \tilde{T}_i$, with the index
$i$ indicating that the field is charged under $SU_i (3)$.

The underlying principle of CSB implies that the low energy
Lagrangian consists of two pieces. The first, ${\cal L}_R$,
preserves a discrete R symmetry and has a supersymmetric R symmetric
minimum of its effective potential. This is the low energy
Lagrangian for the supersymmetric S-matrix of the $\Lambda = 0$
limit. Experience with string theory suggests that it should satisfy
the demands of field theoretic naturalness: every term consistent
with hypothesized symmetries is allowed.  Any term smaller or larger
than would be indicated by Planck scale dimensional analysis should
be explained in terms of an explicit low energy dynamical mechanism.

The second term $\delta {\cal L}$ arises, in a low energy effective
picture, from interactions of a single gravitino with degrees of
freedom on the cosmological horizon in dS space. These DOF {\it do
not have a field theoretic description and we do not yet have a
precise model of them.} We can only list some properties of these
terms, which follow from general principles:

\begin{itemize}

\item They violate the discrete R symmetry.

\item They must give us a low energy effective theory that violates
SUSY, incorporating the relation $m_{3/2} = K \Lambda^{1/4}$.

\item The low energy effective theory must be consistent with a
model of dS space as a system with a finite number of quantum
states. In particular, if the SUSY violating minimum with c.c.
$\Lambda$ is not the absolute minimum of the potential, then the
potential must be {\it Above the Great
Divide}\cite{pyrma}\cite{abj}.

\end{itemize}

The last item implies that the non-gravitational low energy dynamics
must have a ${\it stable} $ SUSY violating ground state\footnote{If
the SUSY violating state is only meta-stable when $m_P \rightarrow
\infty$, and if the difference in energy density between the
meta-stable and "stable" minima is much larger than $\Lambda$, the
gravitational theory is below the Great Divide\cite{abj}\cite{bfl}.}. Results
of Nelson and Seiberg\cite{ns}, when combined with the requirement
that R symmetry is explicitly broken, then imply that the R
violating part of the Lagrangian {\it cannot} satisfy the demands of
naturalness. It must omit terms allowed by all symmetries.  We have
however emphasized that the origin of these terms is novel and
corresponds to nothing in our experience with ordinary string theory
or quantum field models that emerge from quasi-local lattice
dynamics. In models of quantum gravity, the states on horizons,
whether black hole horizons or the cosmological horizon in dS space,
do not have a description in terms of localized bulk degrees of
freedom, obeying the rules of QFT. The R violating terms in the
Lagrangian for local degrees of freedom, are the residuum of
interactions with a large number of horizon states, which decouple
as the dS radius is taken to infinity.  These terms are important,
because they are the origin of supersymmetry breaking. They do not obey the constraints
of naturalness.

The R preserving part of the TeV scale superpotential is the
superpotential of the standard model plus

$$W_R = \sum g_i  S_i \tilde{T}_i T_i  + \sum y_i (T_i^3) +
\tilde{y}_i (\tilde{T}_i)^3  + \sum g_{\mu i} S_i H_u H_d .$$ The R
symmetry, which must have no gauge anomalies, is chosen so that
either $g_1$ or $g_3$ vanishes, as well as one of the pairs
$y_{1,3}$ and $\tilde{y}_{1,3}$. It can also be chosen such that the
coefficients of all baryon and lepton number violating operators of
dimensions $4$ and $5$, apart from neutrino seesaw terms $(H_u
L)^2$, vanish. We require the vanishing of $g_{1\ or\ 3}$ in order
to eliminate SUSY preserving minima. The vanishing of one pair of
the $y$ couplings is introduced in order to have a dark matter
candidate.

The R violating superpotential, coming from interactions with the
horizon, is postulated to be

$$\delta W = W_0 + \sum (m_i T_i \tilde{T}_i ) + \mu_i^2 S_i + \mu H_u
H_d .$$

I'll conclude with a brief list of the properties of the model

\begin{itemize}

\item It has no Supersymmetric minimum at sub-Planckian field
values, and is compatible with an underlying model of dS space with
a finite number of states, incorporating the CSB relation $m_{3/2} =
K \Lambda^{1/4}$.

\item ${\cal L}_R$ has a discrete R symmetry and all R preserving
couplings appear with natural strength. Dimension $4$ and $5$
couplings that violate baryon number are absent, and the only
allowed lepton number violating couplings are the neutrino seesaw
terms $(H_u L)^2$.  The $\mu$ term $H_u H_d$ is also forbidden by R
symmetry. All CP violating angles, apart from the CKM phase, can be
rotated away, and ${\cal L}_R$ has a dangerous axion. The
non-generic terms in $\delta W$ lift the axion. One can argue that
if the origin of CP violation is at energies below the Planck scale,
so that the thermal bath near the horizon is approximately CP
invariant, the CP violating phases in $\delta W$ are very small.
This is a novel solution of the strong CP problem.  The NMSSM
couplings in ${\cal L}_R$ and the explicit $\mu$ term in $\delta W$
give an acceptable Higgs spectrum, without tuning.

\item All couplings are perturbative at the unification scale, and
the model generates a dynamical scale $\Lambda_3 \sim $ a few TeV,
which can explain the origin of gaugino masses.  There is freedom to
separately tune different gaugino masses by using the parameters
$m_i$ in $\delta W$.  The chargino decays promptly in this model, so
that the Fermilab trilepton analysis bounds its mass from below by
$\sim 270$ GeV. By making the parameter $m_3$ reasonably large, we
insure that the gluino is not heavy enough to make dangerous
modifications to the Higgs potential.

\item Dark matter is the pyrma-baryon field $ (T_i )^M_a(T_i )^N_b (T_i
)^K_c \epsilon^{abc} \epsilon_{MNK}$, where $i = 1$ or $3$.  It is
not a thermal relic, but can have the right relic density if an
appropriate pyrma-baryon asymmetry is generated in the early
universe. There will be no dark matter annihilation signals.  The
dark matter particle weighs tens of TeV, and has a magnetic moment.
The magnetic moment leads to an interesting pattern of signals in
terrestrial dark matter detectors, rather different from the signal
for a convential WIMP. The details of this are being worked out.
\end{itemize}
\section{Conclusions}

The theory of holographic space-time seeks to generalize string
theory to situations where the boundaries of space-time are not
asymptotically flat or anti-de Sitter, and the quantum theory does
not have a unique ground state. It builds space-time out of purely
quantum data, the dimensions of Hilbert spaces and common tensor
factors in a net of Hilbert spaces. The topology of a Cauchy surface
is part of the specification of the formalism, and does not change
with time. Space-time geometry is not a fluctuating quantum
variable. Instead the quantum degrees of freedom are quantized
orientations of pixels on the holographic screens of causal
diamonds. Their quantum kinematics is determined by a super-algebra,
whose structure incorporates the quantum remnants of the geometry of
compact dimensions. Compactifications are classified in terms of
possible superalgebras. {\it For finite area holographic screens, compactifications
are finite dimensional unitary representations of a superalgebra,
and have no continuous moduli.  When dS space is modeled as the finite
holoscreen on the cosmological horizon, this automatically leads to a fixing
of moduli. Moduli stabilization is essentially kinematic, and has
nothing to do with an effective potential.}

In the limit that the holographic screen area goes to infinity, and
the screen approaches that of null infinity in Minkowski space, the
pixel variables describe supersymmetric multiplets, including the
gravity multiplet. There is, as yet, no general prescription for
calculating the scattering matrix of this super-Poincare invariant
theory.

The general formalism leads in particular to a completely
non-singular, mathematically complete quantum description of what one
might call the generic Big Bang universe. This is called the dense
black hole fluid (DBHF).  Heuristically, at any time, the particle
horizon volume is completely filled with a single large black hole,
and causally disconnected black holes merge to preserve this
condition as the particle horizon expands. The coarse grained
description of this situation is a flat FRW geometry with equation
of state $p = \rho$. Note that flatness, homogeneity and isotropy
emerge automatically, without any inflation.

An heuristic model of our own universe based on the concept of a low
entropy defect in the DBHF implies that the universe {\it must}
approach an asymptotically dS future, with c.c. determined by
cosmological initial conditions.  dS space is modeled as a quantum
system with a finite number of states, as first envisioned by
Fischler and the present author\cite{tbwf}.

The general formalism of holographic space-time implies that SUSY is
restored as $\Lambda \rightarrow 0$ .  Two arguments, one of which
was reviewed above, suggest that $m_{3/2} = K \Lambda^{1/4}$ .  The
constant $K$ has been argued to be of order $10$, and is related to
the ratio between the unification scale and the Planck scale. When
combined with the desire to explain the apparent unification of
standard model couplings, this low scale of SUSY breaking puts
strong constraints on the effective Lagrangian for particle physics
at TeV energy scales. So far, only a unique class of models has been
found, which can satisfy these constraints.  These are the Pyramid
Schemes, and differ from each other only in the values of a few
parameters. They all have a discrete R symmetry in the $\Lambda = 0$
limit, which is broken by interactions with states on the horizon,
which decouple in this limit. The R violating terms in the effective
Lagrangian do not satisfy the usual laws of naturalness.

The Pyramid Scheme resolves many of the puzzles of low energy
supersymmetric particle physics, some by a novel mechanism. It has
an acceptable level of flavor changing neutral currents and no
dangerous B and L violating operators. It has a novel dark matter
candidate, which carries an approximately conserved $U(1)$ quantum
number and can have the right relic density if an appropriate
asymmetry is generated in the early universe. There are no
annihilation signals. The dark matter candidate is quite heavy and
has a magnetic dipole moment.  Its signals in terrestrial detectors
depend on the target nucleus, and are being worked
out\cite{tbjffst}.   The supersymmetric and strong CP problems, the
$\mu$ problem, and the little hierarchy problem are all resolved by
the non-generic nature of the R violating part of the effective
Lagrangian.

The theory of holographic space-time thus provides a comprehensive
quantum mechanical framework for early and late universe cosmology,
as well as incorporating the surprising connection between the
asymptotic dS nature of the universe and low energy particle
physics. The particle physics implications will be checked, at least
in part, by the LHC. If the theory's predictions are verified, one
would be motivated to attack the unsolved problem of formulating
dynamical equations for holographic space-time.

\section{Acknowledgments}

This research was supported in part by DOE grant number
DE-FG03-92ER40689.


\begin{thebibliography}{19}






\bibitem{holost}  T.~Banks,
  {\it Deriving particle physics from quantum gravity: a plan,}
  arXiv:0909.3223 [hep-th].
 T.~Banks,
  {\it Holographic Space-time from the Big Bang to the de Sitter era,}
  J.\ Phys.\ A  {\bf 42}, 304002 (2009)
  [arXiv:0809.3951 [hep-th]].
T.~Banks,
  {\it II(infinity) factors and M-theory in asymptotically flat space-time,}
  arXiv:hep-th/0607007.
 T.~Banks,
  {\it More thoughts on the quantum theory of stable de Sitter space,}
  arXiv:hep-th/0503066.
   T.~Banks,
  {\it Supersymmetry, the cosmological constant and a theory of quantum  gravity
  in our universe,}
  Gen.\ Rel.\ Grav.\  {\bf 35}, 2075 (2003)
  [arXiv:hep-th/0305206].
 T.~Banks,
  {\it Some Thoughts on the Quantum Theory of de Sitter Space,}
  arXiv:astro-ph/0305037.
\bibitem{bfss}  T.~Banks, W.~Fischler, S.~H.~Shenker and L.~Susskind,
  {\it M theory as a matrix model: A conjecture,}
  Phys.\ Rev.\  D {\bf 55}, 5112 (1997)
  [arXiv:hep-th/9610043].

\bibitem{ns}  A.~E.~Nelson and N.~Seiberg,
  {\it R symmetry breaking versus supersymmetry breaking,}
  Nucl.\ Phys.\  B {\bf 416}, 46 (1994)
  [arXiv:hep-ph/9309299].


\bibitem{susyholo}

 T.~Banks,
  {\it SUSY and the holographic screens,}
  arXiv:hep-th/0305163.
 T.~Banks,
  {\it Breaking SUSY on the horizon,}
  arXiv:hep-th/0206117.
 T.~Banks,
  {\it The phenomenology of cosmological supersymmetry breaking,}
  arXiv:hep-ph/0203066.



\bibitem{holocosm}
 T.~Banks and W.~Fischler,
  {\it The holographic approach to cosmology,}
  arXiv:hep-th/0412097.
 T.~Banks, W.~Fischler and L.~Mannelli,
  {\it Microscopic quantum mechanics of the p = rho universe,}
  Phys.\ Rev.\  D {\bf 71}, 123514 (2005)
  [arXiv:hep-th/0408076].
 T.~Banks and W.~Fischler,
  {\it Holographic cosmology,}
  arXiv:hep-th/0405200.
 T.~Banks and W.~Fischler,
  {\it Holographic cosmology 3.0,}
  Phys.\ Scripta {\bf T117}, 56 (2005)
  [arXiv:hep-th/0310288].
  T.~Banks and W.~Fischler,
  {\it An holographic cosmology,}
  arXiv:hep-th/0111142.
 T.~Banks and W.~Fischler,
  {\it M-theory observables for cosmological space-times,}
  arXiv:hep-th/0102077.

\bibitem{bfm} T.~Banks, B.~Fiol and A.~Morisse,
  {\it Towards a quantum theory of de Sitter space,}
  JHEP {\bf 0612}, 004 (2006)
  [arXiv:hep-th/0609062].

\bibitem{nightmare}  T.~Banks, W.~Fischler and S.~Paban,
  {\it Recurrent nightmares?: Measurement theory in de Sitter space. }
  JHEP {\bf 0212}, 062 (2002)
  [arXiv:hep-th/0210160].
  	
 \bibitem{tbwf}  T.~Banks,
  {\it Cosmological breaking of supersymmetry or little Lambda goes back to  the
  future. II,}
  arXiv:hep-th/0007146.
W.~Fischler, {\it Taking de Sitter Seriously}, talk given at the symposium in honor of G. West, Los Alamos,
June 2000.
\bibitem{abj}  A.~Aguirre, T.~Banks and M.~Johnson,
  {\it Regulating eternal inflation. II: The great divide,}
  JHEP {\bf 0608}, 065 (2006)
  [arXiv:hep-th/0603107].
T.~Banks and M.~Johnson,
  {\it Regulating eternal inflation,}
  arXiv:hep-th/0512141.
\bibitem{bfl} R.~Bousso, B.~Freivogel and M.~Lippert,
  {\it Probabilities in the landscape: The decay of nearly flat space,}
  Phys.\ Rev.\  D {\bf 74}, 046008 (2006)
  [arXiv:hep-th/0603105].

\bibitem{pyramid1}  T.~Banks and J.~F.~Fortin,
  {\it A Pyramid Scheme for Particle Physics,}
  JHEP {\bf 0907}, 046 (2009)
  [arXiv:0901.3578 [hep-ph]].

\bibitem{pyrma} T.~Banks and J.~F.~Fortin,
  {\it Tunneling Constraints on Effective Theories of Stable de Sitter Space,}
  Phys.\ Rev.\  D {\bf 80}, 075002 (2009)
  [arXiv:0906.3714 [hep-th]].

 T.~Banks and J.~F.~Fortin,
  {\it A Pyramid Scheme for Particle Physics,}
  JHEP {\bf 0907}, 046 (2009)
  [arXiv:0901.3578 [hep-ph]].
 T.~Banks, J.~D.~Mason and D.~O'Neil,
  {\it A dark matter candidate with new strong interactions,}
  Phys.\ Rev.\  D {\bf 72}, 043530 (2005)
  [arXiv:hep-ph/0506015].
\bibitem{trin} S.~L.~Glashow,
  {\it Trinification Of All Elementary Particle Forces,}
  CITATION = PRINT-84-0577-BOSTON.


\bibitem{tbjffst} T.~Banks, J-F.~Fortin, S.~Thomas, {\it Fermionic dark matter with electro-magnetic dipole moments}, In preparation.





\end{thebibliography}
\end{document}